\documentstyle [12pt,epsfig]{article}
\def\epsfig#1{}
\begin{document}
\setlength{\baselineskip}{20pt}
\parindent=40 mm
\newcounter{subeq}
\renewcommand{\baselinestretch}{1.0}
\renewcommand{\theequation}{\arabic{equation}\alph{subeq}}
\newcommand{\subeqno}{\stepcounter{subeq} \addtocounter{equation}{-1}}
\newcommand{\alabel}[1]{\addtocounter{equation}{-1}%
                        \refstepcounter{equation}%
                        \label{#1}%
                       }
\newcommand{\subeqres}{\setcounter{subeq}{0}}
\newcommand{\eqnoinc}{\addtocounter{equation}{1}}

\title{\Large\bf  Pion and neutron production by cosmic-ray muons underground}
\author{ Jean Delorme and  Magda Ericson\thanks{ Also at Theory Division, CERN,
CH-1211 Geneva
 23,
Switzerland} \\
{\small Institut de Physique Nucl\'{e}aire et IN2P3, CNRS,
Universit\'{e} Claude Bernard Lyon I,}
{\protect \vspace{-2mm}}
\and {\small 43 Bd. du 11 Novembre,
F-69622 Villeurbanne Cedex, France} \\ \\
\and Torleif Ericson\thanks{Also at
Institute of Theoretical Physics, P.O. Box 805,
S-75105 Uppsala, Sweden } \\
{\small Theory Division, CERN,
CH-1211 Geneva 23, Switzerland } \\ \\
\and Petr Vogel   \\
{\small Physics
Department, California Institute of Technology, Pasadena, CA 91125 } \\ }

\maketitle
\begin{abstract}
{\normalsize
The production of positive pions and neutrons
by cosmic muons at underground sites of various depths
is investigated.
We first test the equivalent photon method in the
particular case of $\Delta$ excitation
by the muon. We find that this method, when it neglects the momentum dependence
of the transverse response, reproduces remarkably well
the theoretical muon cross section.
This success has lead us to apply the method to higher energies,
where it has not been tested. We evaluate in this way the
production of positive pions in
liquid scintillator from known photo-absorption cross sections.
At a shallow depth of 20 meters our estimate
reproduces  the measurement. As for the neutron emission, we include the
obvious sources, such as the giant-resonance excitation, the
quasi-deuteron  process, the quasi-free pion production as well
as neutrons emitted
following pion capture. Our evaluation underestimates
the number of neutrons produced
and finds a too weak dependence on the depth.
This suggests that secondary neutron
production is important at all depths. }
\end{abstract}

\parindent=30pt
\section {\bf  Introduction }
The passage of a high energy charged particle through matter is an
interesting source of electromagnetic nuclear reactions largely occurring
at very small angles.
One can view such processes as produced by the flux of a beam of nearly
 real equivalent photons \cite {WEI34},\cite {WIL35},\cite {DAL57}.
Though these photons are virtual, their kinematics at
small angles of deviation of the charged particle is so close to that of
real photons that it becomes a good approximation to link the
inelastic cross sections for charged particles to the ones
for physical photons with an energy equal to the energy loss. The nuclear
reactions  induced by the small angle scattering of charged particles
are of particular importance to the low count
rate underground experiments, such as searches for neutrino oscillations.
The reason is that cosmic ray muons traverse the experimental area and
their energy loss produces background neutrons and pions.

It is by no means a trivial matter to obtain
reliable estimates for actual yields
of neutrons and pions. The problem is related both to the question of
obtaining a sufficiently reliable estimate for the equivalent photon flux as
well as to the problem of finding appropriate input data.  This is the
aim of the present investigation which has been
stimulated by the rate of low-energy
neutrons observed in an exploration of background sources in a planned
neutrino oscillation experiment.  As a quick estimate has shown, the neutron
yield from the low-energy photo-nuclear giant-resonance excitation,
as considered
e.g. in Ref.~\cite{o'conn}, is much too small to explain the observations,
which suggests rather that meson-producing processes are primarily
responsible. Indeed, negative pions are an
efficient source of neutrons since most of them will slow down and stop
in the surrounding matter.  They are then captured into orbits of pionic
atoms and produce neutron pairs by the quasi-deuteron absorption process
$\pi^-~+~'d' \rightarrow ~2n$ on the central nucleus.  The observation of
correlated neutrons is therefore linked to the yield
of negative pions.  To confirm this mechanism for neutron production we
advocated a study of the yield of charged pions. Negative pions are not
directly observable, but positive
pions are produced at nearly the same rate. Their presence would be an
indication in favor of this mechanism. The yield of positive pions has
since been  directly established
experimentally from the observation of the delayed muon decay from the
 $\pi^+$ stopping without strong interactions \cite{ralph}.

For photons above 300 MeV an efficient way to produce pions is via
the $\Delta$ isobar, which appears as a prominent peak in the
photo-absorption cross sections. In view of the
importance of the isobar region it is interesting to explore the validity
of various approximations to the flux of equivalent photons there.
In this case the isobar excitation by muons can be calculated without
resorting to the equivalent photon method,
and compared to the equivalent photon method in various limits as is done in
Section 2.2, which follow the framework given in Section 2.1. This
provides a guide for the application of this method at higher energies.  In
Section 3 we evaluate
the $\pi^+$ production yield by high energy muons  utilizing the equivalent
photon method. For this we need the photon partial cross sections for a given
multiplicity of $\pi^+$ which, in turn, leads to the discussion of the
assumptions necessary to deduce the yield of positive pions from experimental
data. The somewhat unexpected
result of this study is that pions are produced primarily above
the $\Delta$ isobar region, often by multiple pion production.

\section {\bf  The equivalent Photon Method}
\subsection {General Framework}

We consider a charged lepton of mass $m$ and initial
momentum and energy {\bf P}, $E$
scattering on a stationary target of mass $M$. The final lepton
momentum and energy is {\bf P}', $E'$. The energy and momentum transferred to
the nucleus are, respectively,
\begin{equation}
\nu = E - E'  ~,~  {\bf q} = {\bf P} - {\bf P'}  ~~,
\end {equation}
and $\theta$ is the laboratory scattering angle
between ${\bf P}$ and ${\bf P'}$.

The doubly differential cross section in the laboratory frame
is expressed in terms of the usual structure functions $W_{1,2}$.
Here we adopt the convention used by Close\cite{CLOSE},
where they have the dimension [energy]$^{-1}$.
\begin{equation}
\frac{{\rm d}^2 \sigma}{{\rm d} \Omega {\rm d} E'} =
\frac{2 \alpha^2}{Q^4} \frac{P'}{P}
[ (Q^2 - 2m^2)W_1 + (2EE' - \frac{Q^2}{2})W_2 ] ~,
\end{equation}
where  $Q^2={{\bf q}}^2-\nu^2$.
Now we introduce the transverse and longitudinal responses $R_T$ and
$R_L$ defined by:
\begin{equation}
W_1 = \frac{R_T}{2} ~~;~~ W_2 = \frac{Q^4}{{{\bf q}}^{4}} R_L +
\frac{Q^2}{2 {{\bf q}}^2} R_T ~,
\end{equation}
so that
\begin{equation}
\frac{{\rm d}^2 \sigma}{{\rm d} \Omega {\rm d} E'} =
\alpha^2 \frac{P'}{P} \left [ \frac{4EE' - Q^2}{{{\bf q}}^4} R_L
+ \frac{4EE' + {{\bf q}}^2 + \nu^2 -
4m^2{{\bf q}}^2/Q^2}{2{{\bf q}}^2Q^2} R_T
\right  ] ~.
\label{eq: rose}
\end{equation}
To the order $m^2/E^2 \ll 1$ we have
\begin{eqnarray}
Q^2 \equiv {\bf q}^2 - \nu^2& \equiv& (P-P')^2-\nu ^2+4PP'sin^2\frac{\theta}{2}
\nonumber \\
                       & \simeq&a^2 + 4EE' ( 1 - \frac{m^2}{EE'} -
 \frac{a^2}{2EE'} )\sin^2 \frac{\theta}{2} ~,
\label{eq: kinematics}
\end {eqnarray}
where
\begin{equation}
a^2 = \frac{m^2 \nu^2}{PP'}\simeq \frac{m^2 \nu^2}{EE'} \ll \nu ^2 ~.
\end {equation}
The allowed $Q^2$ values at different incident and transferred energies are
 shown in Fig.~\ref{fig:xsec}.

To make the link with the photo-absorption cross section,
we define the effective number of photons $N(\nu)$ by
\begin{equation}
\frac{{\rm d} \sigma}{{\rm d} E'} = \int {\rm d} \Omega
\frac{{\rm d}^2 \sigma}{{\rm d} \Omega {\rm d} E'}
 = \frac{N(\nu) \sigma_{\gamma} (\nu)}{\nu } ~,
\label{eq: numphot}
\end{equation}
where the photon cross section $\sigma_{\gamma}$ is related to the transverse
response by \cite{SAK69}
\begin{equation}
\sigma_{\gamma} (\nu) = \frac{2\pi^2\alpha}{\nu} R_T(\nu,Q^2 = 0) ~.
\end{equation}
The simplest level of approximation  neglects the
longitudinal part of the response function and also ignores all momentum
dependence of the transverse response, {\it i.e.}, $ R_T(\nu,Q^2) \cong
R_T(\nu,Q^2 = 0)$. In this case the angular integration in Eq.~(\ref{eq:
numphot}) can be performed using
\begin{equation}
{\rm d}Q^2 \equiv \frac{PP'{\rm d} \Omega}{\pi} ~.\label{eq: omq2}
\end{equation}
It leads to the formula:
\begin{equation}
N(\nu) = \frac{\alpha}{\pi}\left [ \frac{E^2+E'^2}{P^2} \ln \frac{EE' + PP' -
m^2}{m\nu} - \frac{(E+E')^2}{2P^2} \ln \frac{(P+P')^2}{(E+E')\nu} -
\frac{P'}{P} \right ]  ~,
\label{eq: DY17}
\end{equation}
which is nothing else than Eq.~(1.7) of Dalitz and Yennie \cite{DAL57}
(with the corrected sign in front of $m^2$).

It is appropriate here to make the link with the commonly used
Weizs\"{a}cker-Williams method \cite{JAC63}. Starting with the Eq.
(\ref{eq: rose}) we make the same assumptions as for the derivation of
Eq. (\ref{eq: DY17}). Moreover, we put $E'/E \approx 1$
(i.e., a small energy loss), expand for
small angles, $Q^2 \approx a^2 + EE'\theta^2$,
${{\bf q}}^2 \approx \nu^2+ EE'\theta^2$, and neglect
${{\bf q}}^2 + \nu^2$ with respect to $4EE'$, obtaining
\begin{equation}
\frac{{\rm d} \sigma}{{\rm d} E'} \approx
\frac{ 2 \pi \alpha^2 R_T(\nu,Q^2 = 0)}{\nu^2} \left [
\ln \frac{ 1 + E^4 \theta_{Max}^2/m^2 \nu^2}
{ 1 + E^2 \theta_{Max}^2/\nu^2} - \frac{\theta_{Max}^2}
{\theta_{Max}^2 + \nu^2 m^2/E^4} \right ] ~,
 \label{eq: stmax}
\end{equation}
where $\theta_{Max}$ is a cut-off scattering angle.
We also assume that $\theta_{Max} \gg m\nu/E^2 = a/E$
and therefore the effective photon number is
\begin{equation}
N(\nu) \simeq \frac{\alpha}{\pi } \left [ 2 \ln \frac{\theta_{Max}E^2}
{m(\nu^2 + \theta_{Max}^2E^2)^{1/2}} - 1 \right ] .
\label{eq: nth}
\end{equation}
This is now the formula that can be compared to the
Weizs\"acker-Williams method \cite{JAC63}:
\begin{equation}
N(\nu) = \frac{ \alpha}{\pi} \frac{c^2}{v^2} \left [
 2 \ln \frac {1.123 p}{m \nu b_{min}} - \frac{v^2}{c^2} \right ] ~.
 \label{eq: WW}
 \end{equation}
Here $v \simeq c$ is the muon velocity, $p$ its momentum,
and $b_{min}$ is the minimum impact parameter. Formula (\ref{eq: WW}) is
essentially equivalent to our Eq. (\ref{eq: nth})
provided we identify $b_{min} = 1/E\theta_{Max}$ for $\theta_{Max} < \nu/E$
and $b_{min} = 1/\nu$ for  $\theta_{Max} > \nu/E$.

Moreover, Eq.~(\ref{eq: nth}) is  identical to Eq.~(1.7) of
Dalitz and Yennie \cite{DAL57} and our Eq. (\ref{eq: DY17}) if we take
 $\theta_{Max} = \nu/E$, and assume that $\nu \ll E$.
In fact, the integral over d$\Omega$ leading to Eq. (\ref{eq: DY17}) gets
the largest contributions from scattering angles
$\theta \le \nu/E$ and saturates at $\theta \simeq \nu/E$
, which justifies the identification $\theta_{Max} = \nu/E$
above, and is compatible with our assumption that that $\theta_{Max} \gg a/E$.
Equations (\ref{eq: nth}) and (\ref{eq: WW}) become therefore equivalent when
we take $b_{min} = 1/\nu$, while Ref.\cite{JAC63} suggests instead the
identification of $b_{min}$ with the larger of the quantities $1/E$ or
$ R$, where $R$ is the nuclear radius. The presence of an undetermined
parameter $b_{min}$ is a serious drawback of
the classical Weisz\"acker-Williams
method, and its absence makes Eq. (\ref{eq: DY17}) attractive.

The real question is whether our assumption that
\begin{equation}
R_T (\nu, Q^2 \leq \nu^2) ~\simeq R_T (\nu, Q^2 = 0)
\end{equation}
is justified. To establish this,
the following section  explores this question for quasi-elastic $\Delta$
excitation for which the photon number can
be calculated exactly, which allows tests  of various approximations.

\subsection {\bf Test of the Method for $\Delta $ Resonance Production}
\subsubsection {\bf Nucleons at rest}

The choice of the $\Delta $ resonance is guided
by two considerations. First, it
is a prominent feature in the photo-absorption  cross section.
Second, in a given model it is tractable
exactly.  It can therefore be used as a test for different approximations to
the equivalent photon method. This study will be done under the assumption that
the $\Delta$ isobar has no width. We also ignore the
quadrupole excitation of the $\Delta$ resonance, i.e., we take the longitudinal
response to vanish. In a first step we omit the Fermi motion of the nucleons.
It is introduced later with no major  change in  the conclusions.
 With these assumptions the transverse nuclear
response per nucleon $R_T$ for $\Delta$ resonance excitation  has  the
 following form, according to Chanfray {\it et al.}~\cite{CHA93}:
\eqnoinc
\begin{eqnarray}
R_T &\approx& \frac{2}{9} G^2 _{M^*}(Q^2) \frac{(M^* - M)^2+Q^2}
{4M^2} \delta( \nu - \nu _{\Delta}-\frac{Q^2}{2M}) \subeqno \alabel{rdelta1} \\
\nonumber \\
& ~=~&\frac{2}{9} G^2 _{M^*}(Q^2) \frac{{\bf q^2}}
{(M^*+M)^2+Q^2} \delta( \nu - \nu _{\Delta}-\frac{Q^2}{2M})), \subeqno
\alabel{rdelta2} \\
\subeqres
\nonumber
\end{eqnarray}
where $\nu_{\Delta}=(M^{*2}-M^2)/2M$, while $M^* ~(M)$ is  the $\Delta$
(nucleon) mass and $G_{M^*}(Q^2)$  the $N\Delta$ transition form factor.
We use $G_{M^*}(0) = \mu^*[(M^*+M)/2M]^2$
with $\mu^*=2\mu_V = 2\times4.71$ n.m., and assume the
usual dipole form for its $Q^2$ dependence with the standard cut-off parameter
$\Lambda = 6m_{\pi }$. The response depends obviously on $Q^2$, not
only because it has magnetic character but also because of the
recoil effect. In addition, there is of course the  $Q^2$ dependence
coming from the vertex form factor.
Substituting for $R_T$ in Eq.~(\ref{eq: rose}) (where we have taken $R_L = 0$)
we can integrate over d$\Omega$, alternatively over d$Q^2$,
using Eq.~(\ref{eq: omq2}).
Note that  since we treat the $\Delta$ as a sharp state,  we cannot use
directly  the relation~(\ref{eq: numphot}) between the photon cross section,
$\sigma_{\gamma}(\nu)$, and the
differential cross section for muons, d$\sigma$/d$E'$,
to define an effective photon number $N(\nu)$ depending on the energy loss.
Indeed, due to recoil effects, the muon cross section
is shifted to a somewhat different  energy with respect to the photon one.
The number of equivalent photons
can be defined only from the energy-integrated cross sections, as follows:
\begin{equation}
N_{\gamma}=\frac{\int {\rm d}E'{\rm d}\sigma (E')/{\rm d}E'}{\int {\rm d}\nu
 \sigma _{\gamma}(\nu)/\nu}
 ~. \label{eq: ngamma}
 \end{equation}
Therefore the total cross section for $\Delta$ resonance excitation is
needed,and it is obtained by integrating again Eq.~(\ref{eq: rose}) over
the final energy $E'$ ({\it i.e.}, over the excitation energy ${\nu}$).
The domain of integration in
the $(Q^2,\nu)$ plane is determined by two requirements. First, the argument of
the $\delta$-function defines the dispersion line
\begin{equation}
\nu = \frac{Q^2}{2M} + \nu_{\Delta} \label{eq: dislim}
\end{equation}
outside of which the response vanishes
(see the full line in Fig.~\ref{fig:xsec}).
Second, the kinematical conditions of muon scattering (\ref{eq: kinematics})
restrict the variation of $Q^2$ between minimum and maximum values
attained at $0^{\circ}$ and $180^{\circ}$. It is then
easily found that one has only access to the part of the ($Q^2, \nu$)
plane lying below the line defined by
\begin{equation}
\nu = \frac{1}{2m^2} \left[ - EQ^2 + P\sqrt{Q^2(Q^2 + 4m^2)}\right].
\label{eq: nulim}
\end{equation}
This is illustrated for  muons of  different energy by the broken curves in
Fig.~\ref{fig:xsec}. They rise  from 0 to the maximum energy loss
$\nu_M = E-m$ which is attained for
$Q^2_M = 2m(E-m)$, and then decreases to 0 again at the maximum possible
transfer $4P^2$. At lepton energies above the $\Delta$ excitation threshold
They intersect the dispersion line of Eq.(\ref{eq: dislim}) twice. The
intersection points define the limits of
the domain of integration along this line and
therefore determine $Q^2_{min}$ and $Q^2_{max}$~. These are given by
the roots of the equation
\begin{equation}
Q^4(\frac{2E}{M} + 1 + \frac{m^2}{M^2}) - 4Q^2\left[ P^2-\frac{\nu_{\Delta}}
{M}(EM + m^2)\right ] + 4m^2\nu_{\Delta}^2 = 0~,   \label{eq: kap}
\end{equation}
which are approximately
\begin{eqnarray}
 Q_{min}^2 & \approx & \frac {m^2 \nu_{\Delta}^2}{
 E(E-\nu_{\Delta })}
 \nonumber \\  Q_{max}^2
 & \approx &
 \frac {2E(E-\nu_{\Delta })M}{(E+M/2)} . \label{eq: klim}
\end{eqnarray}
The value of $Q_{min}^2$ decreases rapidly with
increasing energy $E$ as seen in
Fig.~\ref{fig:xsec}. We find for example the values
392, 13 and 0.13 (MeV/c)$^2$ for $Q^2_{min}$ for the incident energies  E = 2,
10 and 100 GeV, respectively.  The region close to this limit dominates the
integral, since the cross section is sharply peaked at small angles (see
Eq.~(\ref{eq: rose})).
\\
For real photons
the cross section for a $\Delta $ excitation of negligible width is
\begin{equation}
\sigma_{\gamma}(\nu)=\frac{2\pi^2\alpha}{\nu}R_T(Q^2=0,\nu)=
\frac{4\pi^2\alpha}{9}
G_{M^*}^2(Q^2=0)\frac{(M^*-M)}{2M(M^*+M)}\delta(\nu -\nu _{\Delta})
.\label{eq: sigphot}
\end{equation}
with its inverse energy-weighted integral
 \begin{equation}
\int {\rm d}\nu \sigma_{\gamma}(\nu)/\nu =\frac{4\pi^2\alpha}{9}
\frac{G_{M^*}^2(Q^2=0)}{(M^*+M)^2}~.\label{eq: siginv}
\end{equation}
The equivalent photon number  deduced through Eq.~(\ref{eq: ngamma})
from our calculated muon and
photon cross sections is an exact one within the model.
It can be compared with various approximations given by Dalitz and Yennie, in
particular Eq.~(\ref{eq: DY17}) obtained under the assumption of a
momentum independent transverse response.
This expression which is a function of
the energy has to be evaluated at the excitation energy $\nu_{\Delta}$ of the
$\Delta$ isobar by photons.
This comparison is summarized in Table \ref{Tab. 2}  for muons of  different
energies. Surprisingly the agreement between our exact photon number
and the one given by the Dalitz-Yennie formula~(\ref{eq: DY17})  is excellent,
well  beyond what is
expected from the $Q^2$ dependence of the response. In order to elucidate the
origin of this intriguing agreement we have proceeded in several steps. First,
we have suppressed one obvious source of $Q^2$ dependence of the
response in our calculation, namely that of
the magnetic form factor that we now
take to be constant. As shown in Table \ref{Tab. 2}, this has a sizable effect
since this suppression increases the muon cross section by about 30-40 \%,
destroying the previous agreement.

It is then natural to  try to cure this
problem by using instead the formula~(1.8) of Dalitz and Yennie \cite{DAL57},
appropriate for a magnetic transition with a response proportional to ${\bf
q^2}$ and no further momentum dependence:
\begin{equation}
N(\nu) = \frac {\alpha}{\pi} \frac {E^2+E'^2}{P^2} \ln \frac {EE' + PP' -
m^2}{m\nu} . \label{eq: DY18}
\end{equation}
It turns out that this approximation is a poor one. It does not reproduce our
exact result even when we  use a constant form factor
(see Table \ref{Tab. 3}). The
reason is that the response is not only proportional to ${\bf q^2}$,  but there
are additional sources of momentum dependence, as seen in the second form
of the response, Eq.~(\ref{rdelta2}). Their effect is sizable. This is
illustrated  in  Table \ref{Tab. 3}: in line b) we introduce the recoil effect
in the energy conserving $\delta$-function, which produces a momentum
dependence.  In line
c), instead,  we introduce the momentum dependence of the denominator of
Eq.~(\ref{rdelta2}), which acts as a  form  factor. Both have the effect
of reducing the muon cross section by sizable amounts.
Therefore, the expression~(\ref{eq: DY18}),
which in principle should be more appropriate
for a magnetic transition, is in fact a poor
approximation.~Eq.~(\ref{eq: DY17}) (the Dalitz-Yennie 'standard value'), which
assumes a constant response, agrees much better with our result, in particular
 with the one where the variation of the form
 factor is kept. The reason is that the decrease of
the form factor cuts off the large momenta and makes the assumption of a
constant response better which leads to the agreement with
 formula~(\ref{eq: DY17}). We have checked this by
exploring the influence of the form factor on the constant term $(M^*-M)^2$ of
the response in Eq.~(\ref{rdelta1}) and on the $Q^2$ part (see lines
b) and c) of Table~\ref {Tab. 4}). Notice that the first numbers in line a)
reproduce those of line a) of Table~\ref{Tab. 2} with a constant response.
While the form factor influence is only moderate on the first part,
it is quite large in the $Q^2$-term, which is strongly suppressed. This
makes the approximation of a constant response a valid one. A
good agreement  between our value and the formula (\ref{eq: DY17}) is
therefore  understandable, although such a perfect one is a coincidence
depending on the exact value of the cut-off parameter. This is illustrated by
line d) in Table~\ref{Tab. 2} where we have used a larger value
$\Lambda = 8 m_{\pi}$~. The agreement remains good although not quite
perfect. It is difficult to prove the general validity of the Dalitz-Yennie
formula for other production mechanisms.  Consequently, one cannot give
a universal expression for the equivalent photon number, which allows the
derivation of the muon cross section from the photon one with good accuracy.
In the absence of this universal relation and in view of the good
agreement provided by the expression (1.7)  of Ref.\cite{DAL57}
we have thus used it at all energies and,
in particular, also in the multipion range.
The accuracy of such an approximation is obviously difficult to assess.

\subsubsection {\bf Effect of Fermi motion}

We investigate now the influence of the Fermi motion of the nucleons. The
nuclear response in the $(Q^2,\nu)$ plane is no longer restricted by
Eq. (\ref{eq: dislim}) to the line $\nu = Q^2/2M + \nu_{\Delta}$
(dotted line on Fig.~\ref{fig:xsecf}), but now falls in  a band
 between the two curves (the full lines in same figure)
\begin{equation}
\nu_{\pm} = \frac{1}{2M^2} \left[ E_F(Q^2+2M\nu_{\Delta})
\pm p_F\sqrt{4M^2Q^2 +
(Q^2+2M\nu_{\Delta})^2} \right] , \label{eq: nuf}
\end{equation}
where $p_F$ and $E_F$ are the Fermi momentum and energy
($E_F=\sqrt{M^2+p_F^2}$). They meet the vertical axis at $\nu_\pm(0)
= (E_F \pm p_F)\nu_{\Delta}/M$~. Each of them has two intersections with the
kinematical line~(\ref{eq: nulim}) which limits the accessible part of the
$(Q^2, \nu)$ plane. The
integration domain is thus now a surface delineated by four sections of curves.
In the variable $ Q^2$ the absciss$\mbox{\ae }$  of the intersection
points are defined by the roots of equations
\begin{eqnarray}
\lefteqn{Q^4 \left[ \frac{2(EE_F\pm Pp_F)}{M^2} + 1 + \frac{m^2}{M^2} \right]
 -4Q^2 \left[
\frac{(PE_F\pm Ep_F)^2}{M^2}\right.} \nonumber \\
& &\left. - \nu_{\Delta}\frac{(EE_F\mp Pp_F + m^2)}{M} \right] +
4m^2\nu_{\Delta}^2 = 0,
\label{eq: kapf}
\end{eqnarray}
which give back Eq.~(\ref{eq: kap}) in the limit $p_F = 0$.
The corresponding values of $\nu$ are obtained by substitution of the solutions
 in eqs.~(\ref{eq: nulim}) or (\ref{eq: nuf}).

We have made the estimate with a Fermi
momentum of 230 MeV/c, which is a characteristic value for a light nucleus like
 carbon
and using the analytic expressions for the response of a Fermi gas given in
Ref.~\cite{CHA93}.
The resulting effect of Fermi motion on the average photon number $N_\gamma$
is unimportant at lower lepton energies,
and it represents  only a 2.5 \% effect at 15 GeV.
It becomes non negligible in the region of 30 GeV and above. The
vicinity of the singularity makes the integral
sensitive to the detailed behavior of the integrand and thus to Fermi
spreading so that $N_{\gamma} $ deviates  somewhat  from its
free nucleon value. We find, for instance, a decrease of the photon
number by 6, 11, 19 and 27 \% at 20, 30, 50 and 100 GeV respectively.

\section {\bf  Application}

We now turn to the application of
the equivalent photon method to the production
of pions and neutrons by the cosmic ray muons underground.
The muon flux at a depth $d$, $f_d (E)$, normalized according to $\int
f_d(E) {\rm d} E = 1 $, can be  obtained from the
spectrum on surface as given in Ref. \cite{Gaisser}, taking into account
the energy losses of muons penetrating to a given depth.
In Figure \ref{fig:flux} we show the normalized muon flux for
various shallow depths.
The corresponding total fluxes (in units of  muons per horizontal area of
one m$^2$) are 131 for $d=0$,
33 for $d$=20 m, 12 for $d$=50 m, and 3.4 for $d$=100 m.
One sees that the muon spectrum hardens with depth; the average
muon energies are increasing and for  the considered
$d$ values we get $\langle E \rangle$ = 5.4, 10.3, 13.5, and 22.4 GeV.
At the depth of 500 m the flux is reduced to 0.082 muons/m$^2$
and $\langle E \rangle$ = 80 GeV.

By folding the muon flux  $f_d(E)$  with $N({\nu})$, Eq.(\ref{eq: DY17})
or alternatively Eq.(\ref{eq: nth}),
we obtain the dimensionless effective number of equivalent photons
(per muon) of a given energy at a given depth
\begin{equation}
{\cal N}_d(\nu)  ~=~ \int_{m+\nu}^{\infty}  N(\nu) f_d(E) {\rm d}E ~,
\label{eq: ndep}
\end{equation}

In Figure \ref{fig:neff}  we show the effective photon number
${\cal N}_d(\nu)$ at different depths  for photon energies up to 100 GeV.
In preparing the figure we used Eq.(\ref{eq: DY17}); the results
for Eq.(\ref{eq: nth}) with $\theta_{Max}=\nu/E$ are,
naturally, very similar. Notice that
${\cal N}_d(\nu)$ is at first approximatively constant as a function of $\nu$.
The horizontal part turns into a steep decrease  for
$\nu$ somewhat less than  $\langle E \rangle$, where it is cut-off by the
decreasing muon spectrum.
As a function of depth the horizontal part of ${\cal N}_d(\nu)$
increases as ln$(\langle E \rangle)$, as expected. The
photon energy where ${\cal N}_d(\nu)$
bends down increases, however, linearly with
the average muon energy $\langle E \rangle$.
In Figure \ref{fig:neff}
we used Eq.~(\ref{eq: DY17}) for $N({\nu})$ even very close the lower
integration limit in Eq.~(\ref{eq: ndep}),
where its application is somewhat suspect, because the corresponding
scattering angle is not small.
That region contributes relatively little to the integral, however.

In Section 2.2.1 we have shown that Eq. (\ref{eq: DY17}) relates correctly the
muon-nucleus cross section to the photo-nuclear cross section in the $\Delta$
resonance region. Here we are in a position to justify, at least approximately,
its application also in the higher  energy region. According to the standard
treatment \cite{Gaisser,Adair} the muon energy loss by interaction with nuclei
is governed by the formula
\begin{equation}
\frac{{\rm d}E}{{\rm d}X} =   -b_{nuc} E ~,
\end{equation}
where $b_{nuc}$ is approximately independent of  muon energy,
and  equal to $\approx 0.5 \times 10^{-6}$ per g/cm$^2$.
Using Eq. (\ref{eq: DY17}),
the photo-absorption cross section in Fig.\ref{fig:photo} and integrating over
the photon energy, we can calculate $b_{nuc}$. The quantity
$b_{nuc}$ calculated this way is indeed almost independent
of the muon energy for 1 GeV $\le E \le$ 100 GeV
(it is very slowly rising, by a factor of less than two),
and its magnitude agrees with
the empirical value within its uncertainties.

The reaction cross section in a given exit channel, averaged over the muon
spectrum at a given depth $d$ is obtained by analogy to Eq.(\ref{eq: numphot}),
\begin{equation}
\frac{{\rm d} \sigma}{{\rm d} \nu}
 = \frac{{\cal N}_d(\nu)  ~ \sigma_{\gamma} (\nu) ~ br(\nu)}{\nu }  ~,
\label{eq: numeff}
\end{equation}
where $br(\nu)$ is the branching ratio for the corresponding reaction
channel, e.g. $(\gamma,n)$, $(\gamma,\pi^+)$, etc..
(If  more than one particle of a given kind can be produced, the branching
ratio should also reflect the corresponding multiplicity; see  below).
The cross section
(\ref{eq: numeff}) gives the number of neutrons, $\pi^+$, etc. produced
per muon at the depth $d$ in the {\it primary} reaction. As we pointed
out in the introduction, additional particles, in particular neutrons,
can be produced in secondary reactions. Thus, the yields obtained
with Eq.(\ref{eq: numeff}) are really lower limits for the neutron production.
In order to apply the formalism we have to substitute in Eq.(\ref{eq: numeff})
the corresponding experimental
photo-nuclear cross sections, and branching ratios.

The total  photo-absorption  cross section which we used is shown
in Fig.\ref{fig:photo}. Since the $(\gamma,n)$ reaction is the only reaction
channel of interest for photon energies below the pion production threshold,
we use  this partial cross section at these energies, and the
full cross section above it.
The data are from Ref.\cite{Wyc} for the giant-resonance region and
from Ref.\cite{Ahr} above it. The data are smoothed at higher energies and
extrapolated at very high energies.

We first consider the pion production, which can be evaluated more
reliably.  For this we require the branching ratio $br(\nu)$ for $\pi^+$ and
$\pi^-$ in the $\Delta$ resonance region $\nu$ = 160 - 500 MeV, and in
the higher energy region above. In the high energy region the analysis
is simplified by the empirical observation
that the photo-cross section of $\pi^+$ and $\pi^-$ on a  proton is nearly the
same as the corresponding one on a neutron with the opposite charges for the
pions. For example, the cross section for the reaction
$\gamma + p \rightarrow \pi^+ + \pi^- + p$ is approximately equal to the cross
section for $\gamma + n \rightarrow \pi^- + \pi^+ + n$, and similarly for
$\gamma + p \rightarrow \pi^+ + \pi^0 + n$ and
$\gamma + n \rightarrow \pi^- + \pi^0 + p$.
This feature  follows from isospin invariance if
one of the two isospin components is negligible.
This near equality is observed for all pion producing reactions of interest.
This means that we do not need all partial cross sections,
and in particular, that for a carbon target the number of $\pi^+$
produced is the same one as the number of $\pi^-$.

Since for photon energies above 1 GeV multipion production becomes important,
the proper definition of the branching ratio $br(\nu)$ for pions is
\begin{equation}
br(\nu) = \frac{ \sum_n n \sigma(n\pi^+)}{\sigma_{tot}} ~.
\end{equation}
The $\pi^-$ production has the same branching ratio in carbon,
as we  just  pointed out above.

For photon energies below 1 GeV
the reactions $\gamma + p \rightarrow \pi^+ +n$,
$\gamma + p \rightarrow \pi^+ + \pi^0 + n$,
and $\gamma + p \rightarrow \pi^+ + \pi^- + p$
(and the corresponding ones for the neutron targets) are known and are the only
ones of interest. For photon energies between 1 and 10 GeV we were able to find
the data for the proton and deuterium targets
only at a few isolated photon energies \cite{LAN88}.
The corresponding branching ratio appears to be, however,
a smooth and relatively slowly increasing function of energy.
Thus we  interpolated between the known points. Finally, to extend to
even higher photon energies we extrapolated this smooth trend
linearly in the logarithm of the photon energy. It is encouraging
to note that such an extrapolation leads to an average $\pi^+$ multiplicity of
 about 2.5 at the photon energy of
100 GeV, in a good agreement with the measurement of
Enikeev et al. \cite{Enik} for the electromagnetic showers of similar energy
(note, however, that our slope of multiplicity versus energy is less than the
one measured in Ref. \cite{Enik} at
considerably higher energies 100 - 1000 GeV).

Having determined the branching ratio $br(\nu)$ for the nucleon targets, we
assume that in nuclear targets (in our case in carbon) the same $br(\nu)$
is applicable. Thus the quantity $\sum_n n \sigma(n\pi^+)$ is obtained
by multiplying $br(\nu)$ for nucleons by the total photo-absorption
cross section for carbon. The dashed curve in Fig. \ref{fig:photo} shows this
$\sum_n n \sigma(n\pi^+)$. The rather uncertain cross sections and branching
ratios at the highest energies are, fortunately, not very important for our
final results at shallow depths. This uncertainty, however,
prevents us from extending the calculations to deeper sites.

It is now a simple matter to evaluate
the $\pi^+$ production yield by integrating
Eq. (\ref{eq: numeff}) over the virtual photon energy $\nu$. The results are
presented in Table \ref{tab: pion}. We divide there the contributions to the
integral into three energy bins. Column 2 demonstrates
that the $\Delta$ resonance contributes relatively little, even at very shallow
depths. For $d$ = 20 and 100 meters the bulk of the contribution comes
from the region below 10 GeV (column 3), where the multiplicities and cross
sections are reasonably well known. For the largest depth,  ($d = 500$ m)
there is a large contribution from
photons of 10-100 GeV and contributions from
energies beyond 100 GeV, which are not included,  will increase the total cross
section even further.

Since we want to compare our calculated pion yield with the measured one in
a liquid scintillator, we have to take into account
the large hydrogen component
of the scintillator (H/C ratio of 1.89).
The calculation of the pion yield from the
muon scattering on protons proceeds in a complete analogy with the calculation
for carbon described above. The cross sections and branching ratios are
taken from Ref.\cite{LAN88}. and the resulting pion yields are  quoted in Table
\ref{tab: pion} (the entries in parentheses).

In Table \ref{tab: pion} we express
the $\pi^+$ yield in the units in which it is
usually measured, i.e., the number of particles per muon and per (g cm$^{-2}$).
To transform the cross section into such units we have to multiply it by the
number of carbon nuclei per gram of scintillator, $N_C = 4.3 \times 10^{22}$
 g$^{-1}$, and analogously by the number of hydrogen nuclei
 per gram, $N_H = 8.2 \times 10^{22}g^{-1}$.
It is very encouraging to note that our total yield
(i. e. adding the carbon and hydrogen contributions) is $2.95\times 10^{-6}$ at
 $d$ = 20 m in agreement within errors with the yield
 $3.5 \pm 0.7 \times 10^{-6}$
measured at that depth in Ref. \cite{ralph}.

According to our calculation the pion
yield should increase by a factor of three
when going from $d$ = 20 m to 500 m.
The depth dependence has not been measured for pion
production and would constitute a test of the validity of our approach.

The analogous calculation for  the total yield of  neutrons is much more
problematic, in particular at deeper depths corresponding to high energy
photons.  The  reason is the importance of neutron production by secondary
spallation processes.
We do not treat the secondary processes here,
but we comment below on a possible more
complete treatment of the neutron production. First however, since
the global neutron yield is of considerable practical importance
to underground physics  and some data exist\cite{ralph,Enik,Agli},
we give here a crude estimate recalling that the pion production
region, in particular at higher energy,
is the most important one.

A lower estimate is obtained by using the following assumptions
(all assuming a carbon target): a) in the giant-resonance region
region we use the experimental $(\gamma, n)$
cross section); b) in the quasi-deuteron region up to
the pion threshold we take the
branching ratio to be unity,  since mostly both a neutron and a proton are
emitted simultaneously; c) in the region of  pion production we assume that by
quasi-free processes there is always
a primary branching ratio of 1/2 for neutron
emission (equal number of neutrons and protons), while secondary absorption of
negative pions give an additional branching ratio of $1.8\times n_{\pi^-}$,
since stopping $\pi^-$ gives a neutron pair in 80\%
and a neutron-proton pair in 20\% of the cases.
That is almost certainly an underestimate, particularly for large
energy transfers $\nu$ for which a large number of
 hadrons are emitted in the reaction.

Our corresponding  results for the neutron
production underground are summarized in Table \ref{tab: neutron},
again divided into energy transfer bins of obvious physical significance.
Even at very shallow depths the giant-resonance and quasi-deuteron
regions contribute relatively little. As was the case of pion production,
the asymptotic region, with its large uncertainty, appears  not to be crucial
for shallow depths ($d \le$ 100 m), but it gives a very significant fraction
at deeper sites.

As far as comparison with the experiment is concerned
two comments can be made.
At $d$ = 20 m in Ref. \cite{ralph} two values are quoted. The total neutron
yield is (4$ \pm 0.8) \times 10^{-5}  ~n$ per muon and per (g cm$^{-2}$).
However, when the shower component (possibly originating from outside
the detector) is eliminated,
the neutron yield associated with muons alone is about 2
in the same units (and presumably with a similar error bar $\pm 0.8$
as above). Thus, our calculated number of 0.87
underestimates the neutron production by a factor of
two to four, indicating that secondary neutron
production other than by pion capture
is non-negligible even at such a shallow depth.

The depth dependence, as quoted in  Ref.  \cite{Agli} suggests a fivefold
increase at $d =  500$ m  with a yield of about
$2 \times 10^{-4}  ~n$ per muon and per (g cm$^{-2})$. Our calculation
therefore seriously underestimates the neutron yield at that depth,
suggesting that secondary neutron production becomes crucial.

It is beyond the scope of this work to describe the secondary neutron
production, and therefore the total neutron yields, quantitatively.
We can, however, present some qualitative arguments.
For bulk targets of heavy elements,
the neutron multiplicities for hadron-induced
processes at high energies increase linearly
with the incident energy \cite {VAS83}. It is plausible that this
linearity holds also for light elements. Since photons of energy larger than a
few GeV  have characteristic features of hadronic interactions (Vector Meson
Dominance)  we may suppose that this linearity applies also in this
case. We then apply this statement to the rather light elements  of the
detector of Ref. \cite{Agli} for the  part of the photon spectrum
above a certain energy $\nu_0$. Below this energy we keep the previous
approach, in which the secondaries only come from pions. Within this framework
we can determine the unknown proportionality constant  between the
neutron multiplicity and the energy  by fitting the data on neutron production.
 This
 requires a multiplicity  of 4 to 5 neutrons per GeV of photon energy above the
energy $\nu_0$ at which we assume the hadronic approximation to set in.
The first (second) number corresponds to the value of $\nu_0$ of 2 (10) GeV.
Our considerations, besides giving a  lower limit for the neutron yield,
offer a hint of the scenario which might emerge from such a description.

\section {\bf  Conclusions}

In summary we have investigated the production of positive pions and neutrons
by cosmic muons at underground sites of various depths. We have used the
equivalent photon method, the validity of which we have tested in the case of a
$\Delta$ isobaric model. We have made an exact evaluation of both
the muon and photon cross sections  in the framework of
a specific model for $\Delta$ excitation.
We have found that the equivalent photon method
with a Dalitz-Yennie formula for the photon number reproduces remarkably well
 the theoretical cross section for the $\Delta$
excitation by muons, when the
momentum dependence of the transverse response is neglected
as well as the longitudinal response. This agreement is somewhat unexpected
because of the momentum dependence of a response of magnetic type. Our
investigation shows that the agreement is to a large part due to the effect
of the form factors which cut-off the large momenta.
This success has lead us to apply the Dalitz-Yennie formula (\ref{eq: DY17})
also in the region of the response where it has not been tested.

We have then used the method to evaluate the production of positive pions in
liquid scintillator from known photo-absorption cross sections.
This estimate requires
the knowledge of the branching ratio of the pion channel including the
multiplicity. We find that at a depth of 20 meters our estimate
reproduces  the measurement of Ref. \cite{ralph}.

As for the neutron production, we have shown that the
low-energy mechanisms (giant-dipole excitation, quasi-deuteron
production) are unimportant.  Neutrons associated
with pions appear to be the dominant source.  However, using
the information from the neutron production in direct processes
together with neutrons produced from the absorption of stopped negative pions,
our conservative estimates underpredict the experimentally observed
neutron yield. This indicates that there is considerable secondary
neutron production in the target environment. The quantitative
description of such nuclear cascades is a formidable
problem.  Therefore,
the direct pion production is  a much better test of
our understanding of cosmic  muon
interactions,  while  the total neutron
yield has a direct practical interest as a source of
background at underground sites.

\section{{\bf Acknowledgments}}

We are grateful to Prof. J. Ahrens for data on photo-cross sections at high
energy, and to Ralph Hertenberger and Mark Chen for numerous
discussions regarding their experiment.
Part of this work was done while two of us, M. E. and T. E., were visiting
Caltech. We would like to thank Professor Felix Boehm, and Caltech,
for their hospitality.
This work was supported in part by the U.S. Department of Energy under grant
DE-FG03-88ER40397.

\newpage

\begin{table}
\begin{center}
\caption  {Number $N_{\gamma }$ of equivalent photons for $\Delta $
excitation by muons  ($\times 100$). The lines correspond to
\newline
a) $R_T$ = constant, corresponding to the Dalitz-Yennie 'standard value' Ref.
\protect\cite{DAL57} Eq.~(1.7).
\newline
b) $R_T$ = constant \protect$\times~{\bf  q}^2$ corresponding to Dalitz-Yennie
 Eq. (1.8).
 \newline
c) Full expression for $R_T$ with standard cut-off value  \protect$\Lambda ~
=~6~m_{\pi }$.
\newline
d) Full expression for $R_T$; \protect$\Lambda ~=~8~m_{\pi }$
\newline
e) Full expression for $R_T$; $\Lambda ~=~\protect\infty $ (pointlike vertex).}
\label{Tab. 2}
  \medskip
\begin{tabular}{|c||c|c|c|c|c|c|} \hline
$E$  &2&5&10 &20&50&100 \\
(GeV)&&&&&& \\ \hline \hline
a)&0.93&1.44&1.81&2.16&2.61&2.94   \\
b)&2.05&3.11&3.86&4.56&5.46&6.12   \\ \hline
c)&0.90&1.42&1.79&2.15&2.59&2.92  \\  \hline
d)&1.00&1.53&1.91&2.27&2.72&3.05  \\
e)&1.38&2.07&2.52&2.93&3.42&3.77 \\ \hline
\end{tabular}
\end{center}
\end{table}

\vskip 0.4 cm
\begin{table}
\begin{center}
\caption  {\protect  Number $N_{\gamma } $ of equivalent photons for $\Delta $
excitation by muons ($\times 100$). The lines correspond to
\newline
a)  $R_T$ = constant $\times~${\bf  q}$^2$.
\newline
b)  $R_T$ = constant $\times~${\bf
 q}$^2\delta (\nu -\nu _{\Delta }-Q^2/2M)$.
\newline
c)  $R_T$ = constant $\times~{\bf  q}^2/[(M^*+M)^2+Q^2]$. }
\label{Tab. 3}
  \medskip
 \begin{tabular}{|c||c|c|c|c|c|c|} \hline
$E$ &2&5&10 &20&50&100 \\
(GeV) &&&&&& \\ \hline \hline
a)&2.05&3.11&3.86&4.56&5.46&6.12   \\
b) &1.45&2.23&2.79&3.31&3.98&4.48 \\
c) &1.68&2.27&2.65&3.01&3.47&3.87 \\ \hline
\end{tabular}
\end{center}
\end{table}

\vskip 0.4 cm

\begin{table}
\begin{center}
\caption  { The influence of form factor variation on the photon number for
different parts of the response Eq.~(\protect\ref{rdelta1}).
 The upper numbers of each entry are
calculated with a constant form factor (\protect$\Lambda =\infty $);
 the second ones in
parentheses include the form factor variation with a cut-off parameter
 \protect$\Lambda = 6m_{\pi}$.
 \newline
a)  $R_T$ = constant $\times ~G_M^*(Q^2)$; the first line reproduces
Dalitz-Yennie `standard value'.
\newline
b)  $R_T$ = constant $\times ~G_M^*(Q^2)\times~\delta (\nu -
\nu _{\Delta }- Q^2/2M).$
\newline
c)  $R_T$ = constant $\times ~G_M^*(Q^2)\times~Q^2\times ~\delta
(\nu -\nu _{\Delta }-Q^2/2M).$
  Note that adding lines b) and c) gives back lines c) and e) of
Table~\protect\ref{Tab. 2} as it should. }
\label{Tab. 4}
  \medskip
\begin{tabular}{|c||c|c|c|c|c|c|} \hline
$\mu$ energy (GeV)&2&5&10&20&50&100 \\ \hline \hline
a)&0.93 &1.44 & 1.81 &2.16&2.61 &2.94 \\
 & (0.75)& (1.25)& (1.61)& (1.96)& (2.40)& (2.73) \\ \hline
b)&0.84 &1.36 &1.72 &2.08 &2.52 &2.86 \\
& (0.71)& (1.21)& (1.57)& (1.92)& (2.36)& (2.69)\\ \hline
c)&0.54 &0.71 &0.80 &0.85 &0.90 &0.92  \\
& (0.19)&(0.21) &(0.22)& (0.23)& (0.23)& (0.23) \\ \hline
\end{tabular}
\end{center}
\end{table}

\vskip 0.4cm

\begin{table}
\begin{center}
\caption{Summary of the $\pi^+$ production yields in the liquid
scintillator (in units of  $10^{-6} \pi^+$
per muon and per (g cm$^{-2}$)). First entry in each column is for the carbon
component, and the second entry in parentheses is for the hydrogen component}
\medskip
\label{tab: pion}
\begin{tabular}{|c|cc|cc|cc|cc|}
\hline
depth & \multicolumn{2}{c|}{ $\Delta$ res.}
&\multicolumn{2}{c|}{ high energy }
&\multicolumn{2}{c|}{ asymptotic}
 & \multicolumn{2}{c|}{total} \\
  & \multicolumn{2}{c|}{ 0 - 0.5 GeV}
&\multicolumn{2}{c|}{0.5 - 10 GeV}
&\multicolumn{2}{c|}{ 10 - 100 GeV}
&\multicolumn{2}{c|}{} \\
\hline
&C&H&C&H&C&H&C&H\\
\hline
20 m &  0.62&(0.16) & 1.62&(0.34) & 0.18&(0.04) & 2.42&(0.54)  \\
100 m & 0.75&(0.20) & 2.30&(0.48) & 0.61&(0.12) & 3.66&(0.80) \\
500 m & 1.05&(0.27) & 3.79&(0.78) & 2.22&(0.45) &  7.06&(1.50) \\
\hline
\end{tabular}
\end{center}
\end{table}

\vskip 0.4cm

\begin{table}
\begin{center}
\caption{Summary of the primary neutron production yields in carbon
(in units of  $10^{-6}  n$ per muon and per (g cm$^{-2}$)).}
\medskip
\label{tab: neutron}
\begin{tabular}{|c|c|c|c|c|c|c|}
\hline
depth & giant & quasi-deut. & $\Delta$ res. & high energy & asymptotic &
 total \\
  &  0 - 30 MeV & 30 - 150 MeV & 0.15 - 0.5 GeV & 0.5 - 10 GeV & 10 - 100 GeV &
 \\
\hline
20 m & 1.0 & 1.0 & 2.3 & 4.0 & 0.4 & 8.7 \\
100 m & 1.2 & 1.2 & 2.8 & 5.6 & 1.3 & 12.1 \\
 500 m & 1.6 & 1.6 & 3.8 & 9.2 & 4.6  & 20.8 \\
\hline
\end{tabular}
\end{center}
\end{table}

\clearpage

\begin{figure}
\epsfig{file=fige1.eps,height=12.0cm}\vspace*{2cm}
\caption{\protect Boundaries of the allowed domain of $Q^2$ vs.
 $\nu$
for different muon energies (dot-dashed lines).
The solid line represents the region  of
non-vanishing response for $\Delta$ excitation.
The intersections betwen the
 solid and dot-dashed lines
determine the $Q^2_{min}$ and $Q^2_{max}$. }
\label{fig:xsec}
\end{figure}

\begin{figure}
\epsfig{file=fige2.eps,height=12.0cm}\vspace*{2cm}
\caption{\protect Same caption as Fig.~\protect\ref{fig:xsec} with
account of Fermi motion. The region of non-vanishing
response now lies between
the two solid curves (calculated with $p_F$ = 230 MeV/c)
which come to coincide
with the dispersion line~(\protect\ref{eq: dislim}) in the
limit $p_F$ = 0
 (dotted curve). There are four intersections betwen the
 solid and dot-dashed
 lines which determine the bounds of the integration region. }
\label{fig:xsecf}
\end{figure}

\begin{figure}
\epsfig{file=fige3.eps,height=12.0cm}\vspace*{2cm}
\caption{\protect  Normalized muon flux at
different depths.}
\label{fig:flux}
\end{figure}

\begin{figure}
\epsfig{file=fige4.eps,height=12.0cm}\vspace*{2cm}
\caption{\protect Effective photon number at
different depths according to the dimensionless
definition in Eq. (26), per muon.}
\label{fig:neff}
\end{figure}

\begin{figure}
\epsfig{file=fige5.eps,height=12.0cm}\vspace*{2cm}
\caption{\protect   Total photoabsorption cross section for
carbon (full line) and the weighted $\pi^+$ photoproduction cross section
$\sum_n n \sigma(n\pi^+)$ (dashed line). }
\label{fig:photo}
\end{figure}

\end{document}